

\magnification = 1200

\font\titlerm = cmr10 scaled\magstep 4
\font\titlerms = cmr7 scaled\magstep 4
\font\titlermss = cmr5 scaled\magstep 4
\font\titlei = cmmi10 scaled\magstep 4
\font\titleis = cmmi7 scaled\magstep 4
\font\titleiss = cmmi5 scaled\magstep 4
\font\titlesy = cmsy10 scaled\magstep 4
\font\titlesys = cmsy7 scaled\magstep 4
\font\titlesyss = cmsy5 scaled\magstep 4
\font\titleit = cmti10 scaled\magstep 4

\def\titlefont{\def\rm{\fam0\titlerm}
\def\it{\fam\itfam\titleit}
\textfont0 = \titlerm
\scriptfont0 = \titlerms
\scriptscriptfont0 = \titlermss
\textfont1 = \titlei
\scriptfont1 = \titleis
\scriptscriptfont1 = \titleiss
\textfont2 = \titlesy
\scriptfont2 = \titlesys
\scriptscriptfont2 = \titlesyss
\textfont\itfam = \titleit
\rm}

\def\sectionfont{\def\rm{\fam0\tenrm}
\def\it{\fam\itfam\tenit}
\def\bf{\fam\bffam\tenbf}
\textfont0 = \tenrm
\scriptfont0 = \sevenrm
\scriptscriptfont0 = \fiverm
\textfont1 = \teni
\scriptfont1 = \seveni  \scriptscriptfont1=\fivei
\textfont2 = \tensy
\scriptfont2 = \sevensy
\scriptscriptfont2 = \fivesy
\textfont\itfam = \tenit
\textfont\bffam = \tenbf
\rm}

\font\teenyfont = cmr5

\global\baselineskip = 1.2\baselineskip
\global\parskip = 4pt plus 0.3pt
\global\nulldelimiterspace = 0pt

\predisplaypenalty 1000


\def\endignore{}
\def\ignore #1\endignore{}

\newcount\dflag
\dflag = 0


\def\monthname{\ifcase\month
\or Jan \or Feb \or Mar \or Apr \or May \or June%
\or July \or Aug \or Sept \or Oct \or Nov \or Dec
\fi}




\def\endid{}
\def\id#1\endid{\number\day\ \monthname \number\year
\hfill #1}

\def\endtitle{}
\def\title#1\endtitle{\vskip.05in\titlefont
\global\baselineskip = 2\baselineskip
#1\vskip.3in
\baselineskip = 0.5\baselineskip\sectionfont}

\def\lblfoot{This work was supported by the Director, Office of Energy
Research, Office of High Energy and Nuclear Physics, Division of High
Energy Physics of the U.S. Department of Energy under Contract
DE-AC03-76SF00098.}

\def\endauthors{}
\def\authors#1\endauthors{
#1\if\dflag = 0
\footnote{}{\noindent\lblfoot}\fi}

\def\endabstract{}
\def\abstract#1\endabstract{\vskip .2in%
\centerline{\sectionfont\bf Abstract}%
\vskip .1in%
\noindent#1%
\ifnum\dflag = 0
\footline = {\hfil}\pageno = 0
\vfill\eject
\pageno = 1\footline{\centerline{\sectionfont\folio}}
\fi\ifnum\dflag = 2
\footline = {\hfil}\pageno = 0
\vfill\eject
\fi}


\newcount\nsection
\newcount\nsubsection

\def\section#1{\global\advance\nsection by 1
\global\nsubsection = 0
\bigskip\noindent
\centerline{\sectionfont\bf\number\nsection.\ #1}
\nobreak\medskip\sectionfont\nobreak}

\def\subsection#1{\global\advance\nsubsection by 1
\bigskip\noindent
\centerline{\sectionfont \it \number\nsection.\number\nsubsection.\ #1}
\nobreak\smallskip\rm\nobreak}

\def\appendix#1#2{\bigskip\noindent%
\sectionfont \bf Appendix #1.\ #2
\nobreak\medskip\rm\nobreak}


\newcount\nref
\global\nref = 1

\def\ref#1#2{\xdef #1{[\number\nref]}
\ifnum\nref = 1\global\xdef\therefs{\noindent[\number\nref] #2\ }
\else
\global\xdef\oldrefs{\therefs}
\global\xdef\therefs{\oldrefs\vskip.1in\noindent[\number\nref] #2\ }%
\fi%
\global\advance\nref by 1
}

\def\listrefs{\vfill\eject\section{References}\therefs}


\newcount\nfig
\global\nfig = 1

\def\fg#1\efig{\vskip .5in\noindent Fig.\ \number\nfig:\ #1%
\global\advance\nfig by 1}


\newcount\cflag
\newcount\nequation
\global\nequation = 1
\def\eqlabel{(1)}

\def\nexteqno{\ifnum\cflag = 0
\global\advance\nequation by 1
\fi
\global\cflag = 0
\xdef\eqlabel{(\number\nequation)}}

\def\lasteqno{\global\advance\nequation by -1
\xdef\eqlabel{(\number\nequation)}}

\def\label#1{\xdef #1{(\number\nequation)}
\ifnum\dflag = 1
{\escapechar = -1
\xdef\draftname{\teenyfont\string#1}}
\fi}

\def\clabel#1#2{\xdef\eqlabel{(\number\nequation #2)}
\global\cflag = 1
\xdef #1{\eqlabel}
\ifnum\dflag = 1
{\escapechar = -1
\xdef\draftname{\string#1}}
\fi}

\def\cclabel#1#2{\xdef\eqlabel{#2)}
\global\cflag = 1
\xdef #1{\eqlabel}
\ifnum\dflag = 1
{\escapechar = -1
\xdef\draftname{\string#1}}
\fi}


\def\eeq{}

\def\eqnn #1\eeq{$$ #1 $$}

\def\eq #1\eeq{\xdef\draftname{\ }
$$ #1
\eqno{\eqlabel \rlap{\ \draftname}} $$
\nexteqno}



\def\eol{& \eqlabel \rlap{\ \draftname} \crcr
\nexteqno
\xdef\draftname{\ }}

\def\eeol{& \eqlabel \rlap{\ \draftname}
\nexteqno
\xdef\draftname{\ }}

\def\eolnn{\cr
\global\cflag = 0
\xdef\draftname{\ }}


\def\eqa #1\eeq{\xdef\draftname{\ }
$$ \eqalignno{ #1 } $$
\global\cflag = 0}


\def\etal{{\it et.\ al.\/}}

\def\myinstitution{
    \centerline{\it Theoretical Physics Group}
    \centerline{\it Lawrence Berkeley Laboratory}
    \centerline{\it 1 Cyclotron Road}
    \centerline{\it Berkeley, California 94720}
}


\def\jref#1#2#3#4{{\it #1} {\bf #2}, #3 (#4)}

\def\HPA#1#2#3{\jref{Helv.\ Phys.\ Act.}{#1}{#2}{#3}}

\def\NPB#1#2#3{\jref{Nucl.\ Phys.}{B#1}{#2}{#3}}
\def\PA#1#2#3{\jref{Physica}{#1A}{#2}{#3}}
\def\PLB#1#2#3{\jref{Phys.\ Lett.}{#1B}{#2}{#3}}

\def\PRD#1#2#3{\jref{Phys.\ Rev.}{D#1}{#2}{#3}}

\def\PRV#1#2#3{\jref{Phys.\ Rev.}{#1}{#2}{#3}}


\def\goto{\mathop{\rightarrow}}


\def\frac#1#2{{{#1} \over {#2}}\,}  


\def\Dsl{\hbox{/\kern-.6000em\rm D}} 



\def\scr#1{{\cal #1}}

\def\mybar#1{\kern 0.8pt\overline{\kern -0.8pt#1\kern -0.8pt}\kern 0.8pt}
\def\sla#1{\raise.15ex\hbox{$/$}\kern-.57em #1}
\def\Sla#1{\kern.15em\raise.15ex\hbox{$/$}\kern-.72em #1}

\def\roughly#1{\mathrel{\raise.3ex\hbox{$#1$\kern-.75em%
    \lower1ex\hbox{$\sim$}}}}
\def\lsim{\roughly<}


\def\tr{\mathop{\rm tr}}


\def\bra#1{\langle #1 |}
\def\ket#1{| #1 \rangle}



\def\al{\alpha}
\def\del{\delta}
\def\Del{\Delta}

\def\Lam{\Lambda}

\def\Sig{\Sigma}

\def\CPT{\raise.45ex\hbox{$\chi$}PT}

\def\hc{{\rm h.c.}}


\def\MeV{{\rm \ MeV}}

\hyphenation{ba-ry-on ba-ry-ons}
\def\chisim{$SU(3)_L \times SU(3)_R$}


\id
LBL-33993
\endid
\rightline{CfPA-TH-93-09}

\title
\centerline{SU(3) vs. SU(3) x SU(3) Breaking in}
\centerline{Weak Hyperon Decays}
\endtitle

\authors
\centerline{Markus A. Luty}
\footnote{}{\hskip-.26in\lblfoot}
\vskip .1in
\myinstitution
\vskip .15in
\centerline{Martin White}
\vskip .1in
\centerline{{\it Center for Particle Astrophysics}}
\centerline{{\it 301 Le Conte Hall}}
\centerline{{\it University of California}}
\centerline{{\it Berkeley, CA 94720}}
\endauthors

\abstract
We consider the predictions of chiral perturbation theory for $SU(3)$
breaking in weak semileptonic and $s$-wave nonleptonic hyperon decays.
By defining an expansion sensitive only to $SU(3)$ breaking, we show that
the leading corrections give rise to moderate corrections to $SU(3)$
relations ($\lsim 20\%$), even though the {\it chiral} symmetry
\chisim\ appears to be rather badly broken.
This explains why $SU(3)$ fits to weak hyperon decays work well even
though chiral-symmetry breaking corrections are large.
Applying these $SU(3)$-breaking corrections to the analysis of the EMC
data, we find that the predicted value of
$\bra p\mybar s\gamma_\mu\gamma_5 s\ket p$
is reduced by $\simeq 35\%$, suggesting that the ``EMC effect'' may be
less striking than commonly thought.
\endabstract


\ref\JMax{E. Jenkins and A. V. Manohar, \PLB{255}{558}{1991}.}

\ref\Jnon{E. Jenkins, \NPB{375}{561}{1992}.}

\ref\heavy{H. Georgi, \PLB{240}{447}{1990};
T. Mannel, W. Roberts, and Z. Ryzak, \NPB{368}{204}{1992}.}

\ref\Wise{J. Bijnens, H. Sonoda, and M. Wise, \NPB{261}{185}{1985}.}

\ref\JMdec{E. Jenkins and A. V. Manohar, \PLB{259}{353}{1991}.}

\ref\krause{A. Krause, \HPA{63}{3}{1990}.}

\ref\ourvec{J. Anderson and M. A. Luty, LBL preprint LBL-33435, to
appear in {\it Phys.\ Rev.} {\bf D}.}

\ref\PDG{Particle Data Group, \PRD{45}{S1}{1992}.}

\ref\effL{J. Schwinger, \PLB{24}{473}{1967};
S. Coleman, J. Wess, and B. Zumino, \PRV{117}{2239}{1969};
C. G. Callan, S. Coleman, J. Wess, and B. Zumino,
\PRV{117}{2247}{1969};
S. Weinberg, \PA{96}{327}{1979}.}

\ref\EMC{J. Ashman \etal, \PLB{206}{364}{1988};
\NPB{328}{1}{1989}.}

\ref\muchado{For a theoretical overview and references to the original
literature, see R. L. Jaffe and A. V. Manohar, \NPB{337}{509}{1990}.}


\section{Introduction}

In this paper, we consider corrections to the $SU(3)$ predictions for
weak semileptonic and $s$-wave nonleptonic hyperon decay rates.
The $SU(3)$ predictions are valid in the limit where $m_u = m_d = m_s$
(we neglect electromagnetism), and experimentally they work to better than
$20\%$.
This remarkable agreement is certainly not due to the fact that the
quark masses are nearly equal;
if they were, the $\pi^0$ and $\eta$ would be nearly degenerate in mass,
while we know that $m_\eta / m_\pi \simeq 4$.
Understanding why some $SU(3)$ predictions work well while others fail
completely has been a theoretical challenge since the discovery of
these relations.

To make progress on this question it is clearly necessary to have a
systematic framework to study deviations from $SU(3)$ symmetry.
Chiral perturbation theory provides such a framework, giving a rigorous
expansion around the {\it chiral} limit: $m_u, m_d, m_s \goto 0$.
In the chiral limit, the octet mesons $\pi$, $K$, and $\eta$ are massless
Nambu--Goldstone bosons whose couplings are constrained by the
low-energy theorems of spontaneous symmetry breaking.
These theorems can be encoded in an effective lagrangian with a
non-linearly realized \chisim\ symmetry.
The lowest-order predictions of chiral perturbation theory embody the
$SU(3)$ predictions, and deviations from $SU(3)$ symmetry relations can be
studied by considering corrections to the chiral limit.

Chiral perturbation theory for baryons was recently reformulated by Jenkins
and Manohar using an effective lagrangian in which the baryons are treated
as heavy fields \JMax.
These authors computed the $O(m_s \ln m_s)$ corrections to the
hyperon weak decay form factors \JMax\Jnon\ and found that corrections to
the lowest-order predictions were $\sim 100\%$.
The logarithmically-enhanced corrections are not expected to dominate
the uncalculable $O(m_s)$ contributions in the real world.
However, the large size of the logarithmically-enhanced corrections does
suggest that chiral perturbation theory is breaking down for these
processes, and makes the success of the lowest-order predictions difficult
to understand.
Also puzzling is that the ``corrected'' predictions still fit the data
well, at the price of large shifts in the values of the couplings
which define the chiral expansion.
For example, in ref.\ \JMax, the values to the axial-vector
form factors including the corrections were found to give
$D = 0.56$, $F = 0.33$, while their lowest-order fit gives
$D = 0.80$, $F = 0.50$.

The authors of ref.\ \JMax\ propose that the breakdown of chiral
perturbation theory for baryons coupled to mesons is due to the presence
of the nearby decuplet states \JMdec.
They find that including decuplet intermediate states reduces the size
of the logarithmically-enhanced corrections, but they still require
large shifts parameters to accommodate the data.
We will not consider this point of view here.

In this paper, we propose a well-motivated and well-defined resummation of
the chiral expansion which is sensitive only to $SU(3)$ breaking.
We compute the logarithmically-enhanced contributions to the weak decay
form factors in this expansion, and find that all corrections are
$\lsim 20\%$.
We conclude that there is no reason to believe that this $SU(3)$ expansion
is breaking down, even though the chiral expansion does not seem
to work well.
This is somewhat surprising, since both expansions are controlled by $m_s$
in the limit $m_s \gg m_u, m_d$.
Our conclusion is also supported by the fact that predictions for the
$p$-wave nonleptonic decays, which follow from chiral symmetry
but not from $SU(3)$ alone, do not work well.

We also apply our results to consider the effects of $SU(3)$ breaking on
the interpretation of the EMC effect.
We find that $SU(3)$ breaking reduces the predicted value of
$\bra p\mybar s\gamma_\mu\gamma_5 s\ket p$ by $35\%$, reducing the
size of the ``EMC effect.''

The plan of this paper is as follows.
In section 2, we briefly review the effective lagrangian formalism we will
use to carry out our computations.
In section 3, we discuss the computation of the semileptonic decay rates.
In section 4, we apply our results to the EMC data.
In section 5, we discuss the computation of the nonleptonic decay rates.
Section 6 contains our conclusions.

\section{The Effective Lagrangian}

In this section, we briefly review the effective lagrangian we use
to carry out the computation.
The notation and conventions we use are the same as those of ref.\
\ourvec.
We briefly review the formalism here for completeness.
The reader familiar with this formalism is urged to skip to section 3.

\subsection{Mesons}

The field
\eq
\xi(x) = e^{i\Pi(x) / f},
\eeq
is taken to transform under $SU(3)_L \times SU(3)_R$ as
\eq
\xi \mapsto L \xi U^\dagger = U \xi R^\dagger,
\eeq
where this equation implicitly defines $U$ as a function of $L$, $R$,
and $\xi$.
The meson fields are
\eq
\Pi = \frac 1{\sqrt 2}
\pmatrix{\frac 1{\sqrt 2}\pi^0 + \frac 1{\sqrt 6}\eta &
\pi^+ & K^+ \cr
\pi^- & -\frac 1{\sqrt 2} \pi^0 + \frac 1{\sqrt 6}\eta &
K^0 \cr
K^- & {\mybar K}^0 & -\frac 2{\sqrt 6} \eta \cr}.
\eeq

Since we are interested in matrix elements of the vector- and
axial-vector Noether currents, we add source terms
\eq
\delta\scr L = \scr V^\mu J_\mu^V + \scr A^\mu J_\mu^A
\eeq
by defining the covariant derivatives
\eq
D_\mu \xi \equiv \partial_\mu \xi - i\ell_\mu \xi,
\qquad D_\mu \xi^\dagger \equiv \partial_\mu \xi^\dagger
- ir_\mu \xi^\dagger.
\eeq
(Note that $(D_\mu\xi)^\dagger \ne D_\mu\xi^\dagger$.)
Here,
\eq
r_\mu = \scr V_\mu + \scr A_\mu,
\qquad \ell_\mu = \scr V_\mu - \scr A_\mu.
\eeq
The effective lagrangian is most conveniently written in terms of
\eq
V_\mu \equiv \frac i2\left(\xi D_\mu \xi^\dagger
+ \xi^\dagger D_\mu \xi\right), \qquad
A_\mu \equiv \frac i2\left(\xi D_\mu \xi^\dagger
-\xi^\dagger D_\mu \xi\right),
\eeq
which transform under {\it local} \chisim\ as
\eq
V_\mu \mapsto U V_\mu U^\dagger + iU\partial_\mu U^\dagger, \qquad
A_\mu \mapsto U A_\mu U^\dagger.
\eeq
The covariant derivative
\eq
\label\covder
\nabla_\mu A_\nu \equiv \partial_\mu A_\nu - i [V_\mu, A_\nu],
\eeq
transforms under local \chisim\ as
\eq
\nabla_\mu A_\nu \mapsto U\nabla_\mu A_\nu U^\dagger.
\eeq

The chiral symmetry is broken explicitly by the quark masses.
(We neglect the effects of electromagnetism in this paper.)
We will ignore isospin breaking, so that the quark mass matrix is
taken to be
\eq
\label\massmat
M_q = \pmatrix{\hat m &&\cr &\hat m&\cr &&m_s\cr}.
\eeq
It is convenient to define the even- and odd-parity fields
\eqa
M &\equiv \frac 12\!\left( \xi^\dagger M_q \xi^\dagger + \hc\right)
\;\mapsto U\!M U^\dagger, \eol
P &\equiv \frac 1{2i}\!\left( \xi^\dagger M_q \xi^\dagger -
\hc\right)
\mapsto U\!P U^\dagger. \eeol
\eeq

The simple transformation rules of the fields defined above makes it easy
to write down the effective lagrangian.
For example, the leading terms can be written
\eq
\scr L_0 = f^2 \tr(A^\mu A_\mu) + af^3 \tr M.
\eeq

\subsection{Baryons}

Because we are interested in processes with characteristic energy much
smaller than baryon masses, the baryons may be treated as heavy
particles \heavy\JMax.
The basic idea is to write the baryon momentum as $P=Mv+k$,
where $M$ is the common baryon mass in the $SU(3)$ limit and $v$ is chosen
so that all of the components of the residual momentum $k$ are small
compared to hadronic scales, $\Lam$, for the process of interest.
The effective lagrangian is then labelled by $v$ and is written in terms
of fields $B$ satisfying the positive energy condition $\sla v B = B$,
and whose momentum modes are the residual momenta of the baryons.
This explicitly removes $M$ as a kinematic scale in the problem.

The octet baryon fields $B$ transform under \chisim\ as
\eq
B \mapsto U\!B U^\dagger.
\eeq
Explicitly, we have
\eq
B = \pmatrix{
\frac 1{\sqrt 2} \Sigma^0 + \frac 1{\sqrt 6}\Lambda &
\Sigma^+ & p \cr
\Sigma^- &
-\frac 1{\sqrt 2}\Sigma^0 + \frac 1{\sqrt 2}\Lambda & n \cr
\Xi^- & \Xi^0 & -\frac 2{\sqrt 6}\Lambda \cr}.
\eeq
The lowest order terms in the effective lagrangian involving baryon
fields are
\eq
\label\Beff
\eqalign{
\scr L &= \tr\left(\mybar B iv\cdot\nabla B\right)
+ 2D \tr\left( \mybar B s^\mu \{ A_\mu, B \} \right)
+ 2F \tr\left( \mybar B s^\mu [ A_\mu, B ] \right) \cr
&\qquad + \sigma \tr\left(M \right) \tr\left( \mybar B B \right)
+ b_D \tr\left( \mybar B \{ M, B \} \right)
+ b_F \tr\left( \mybar B [ M, B ] \right), \cr}
\eeq
where the spin matrix is given by
\eq
s^\mu \equiv \frac 12 (\gamma^\mu - \sla v v^\mu) \gamma_5,
\eeq
and the covariant derivative acts on $B$ as in eq.\ \covder.

\section{Semileptonic Decays}

In this section, we consider the $\Delta S = 1$ semileptonic decays
of hyperons.  These decays are governed by the form factors
\eqa
\bra{B_a} J^{V}_{\mu c}(0) \ket{B_b} &= \mybar u(p_a) \biggl[
f_1^{abc}(q^2) \gamma_\mu
+ \frac{if_2^{abc}(q^2)}{M_a + M_b} \sigma_{\mu\nu} q^\nu
+ \frac{if_3^{abc}(q^2)}{M_a + M_b} q_\mu \biggr] u(p_b), \eol
\bra{B_a} J^{A}_{\mu c}(0) \ket{B_b} &= \mybar u(p_a) \biggl[
g_1^{abc}(q^2) \gamma_\mu \gamma_5
+ \frac{ig_2^{abc}(q^2)}{M_a + M_b} \sigma_{\mu\nu} \gamma_5 q^\nu
+ \frac{ig_3^{abc}(q^2)}{M_a + M_b} \gamma_5 q_\mu \biggr] u(p_b),\ \eeol
\eeq
where $q \equiv p_a - p_b$.
In the $SU(3)$ limit $m_u = m_d = m_s$, the form factors at zero momentum
transfer are determined in terms of two parameters, $D$ and $F$:
$f_2(0) = f_3(0) = g_2(0) = g_3(0) = 0$,
the $f_1(0)$ are $SU(3)$ Clebsch--Gordan coefficients, and the $g_1(0)$
are simple linear combinations of $D$ and $F$ (see below).
We consider the form factors at zero momentum transfer because the masses of
the baryon octet become degenerate in the $SU(3)$ limit, so the $q^2$
dependence of the form factors is higher order in the $SU(3)$ expansion.

We will study deviations from the $SU(3)$ limit using chiral perturbation
theory.
The contribution of the form factors $f_3$ and $g_3$ is suppressed by
the electron mass, and can be safely neglected.
The corrections to $f_1$ and the values of $f_2$ and $g_2$ are $O(m_s)$
and are not calculable in chiral perturbation theory.
The corrections to $f_1$ are $O(m_s)$ and are calculable due to the
Ademollo--Gatto theorem; numerically, they are $\lsim 5\%$ \krause\ourvec.
The corrections to $g_1$ are $O(m_s \ln m_s)$, and are therefore formally
the largest corrections in the chiral expansion.
We therefore focus on $g_1$ for the remainder of this section.
In ref.\ \JMax, these corrections were computed, and were found to be
$\sim 100\%$.\footnote{$^*$}
{An earlier calculation \Wise\ which found smaller corrections is
incorrect.}

Aside from the distinction between $SU(3)$ and chiral symmetry breaking,
our calculation differs from that of ref.\ \JMax\ only in that we keep
$m_\pi \ne 0$.
The $\pi$ corrections are expected to be only $\sim 20\%$ of the $K$ and
$\eta$ corrections, but setting $m_\pi = 0$ systematically {\it increases}
the amount of predicted $SU(3)$ violation.

We write
\eq
g_1^{abc}(0) = \al_{ab}^c + \frac 1{16\pi^2 f^2} \beta_{ab}^c,
\eeq
where the lowest-order results are
\eq
\al_{ab}^c = D d_{ab}^c + F f_{ab}^c,
\eeq
where $d_{ab}^c$ and $f_{ab}^c$ are the symmetric and antisymmetric
structure constants of $SU(3)$, respectively.
Specifically,
\eq
\label\loword
\eqalign{
\al_{pn}^{1 + i2} &= D + F, \cr
\al_{\Lam\Sig^-}^{1 + i2} &= \frac 2{\sqrt 6} D, \cr
\al_{p\Lam}^{4 + i5} &= -\frac 1{\sqrt 6} (D + 3F), \cr
\al_{\Lam\Xi^-}^{4 + i5} &= -\frac 1{\sqrt 6} (D - 3F), \cr
\al_{n\Sig^-}^{4 + i5} &= D - F, \cr
\al_{\Sig^0\Xi^-}^{4 + i5} &= \sqrt 2\al_{\Sig^+\Xi^0}^{4 + i5}
= \frac 1{\sqrt 2} (D + F). \cr}
\eeq
The leading chiral corrections are
\def\fpi{\,m_\pi^2 \ln\frac{m_\pi^2}{\mu^2}}
\def\fk{\,m_K^2 \ln\frac{m_K^2}{\mu^2}}
\def\feta{\,m_\eta^2 \ln\frac{m_\eta^2}{\mu^2}}
\eqa
\label\firstcorr
\beta_{pn}^{1 + i2} &=
-(D + F)(2 D^2 + 4 D F + 2 F^2 + 1) \fpi\eolnn
&\qquad - \frac 16 (13 D^3 - D^2 F + 3 D + 3 D F^2 + 3 F + 33 F^3)
\fk\eolnn
&\qquad - \frac 13 (D + F) (D - 3F)^2 \feta, \eol
\beta_{\Lam\Sig^-}^{1 + i2} &=
-\frac 2{3\sqrt 6} D (7 D^2 + 3 F^2 + 3) \fpi\eolnn
&\qquad - \frac 1{\sqrt 6} D (3 D^2 + 13 F^2 + 1) \fk\eolnn
&\qquad - \frac 4{3\sqrt 6} D^3 \feta, \eol
%
%
\beta_{p\Lam}^{4 + i5} &=
\frac 3{8\sqrt 6}
(3 D^3 + 27 D^2 F + D + 25 D F^2 + 3 F + 9 F^3) \fpi\eolnn
&\qquad +\frac 1{12 \sqrt 6}
(31 D^3 + 15 D^2 F + 9 D + 9 D F^2 + 27 F + 297 F^3) \fk\eolnn
&\qquad + \frac 1{24 \sqrt 6}
(D + 3 F)(19 D^2 - 30 D F + 27 F^2 + 9) \feta, \eol
\beta_{\Lam\Xi^-}^{4 + i5} &=
\frac 3{8 \sqrt 6}(3 D^3 - 27 D^2 F + D + 25 D F^2 - 3 F - 9 F^3)
\fpi\eolnn
&\qquad + \frac 1{12\sqrt 6}
(31 D^3 - 15 D^2 F + 9 D + 9 D F^2 - 27 F - 297 F^3) \fk\eolnn
&\qquad + \frac 1{24\sqrt 6}
(D - 3F)(19 D^2 + 30 D F + 27 F^2 + 9)\feta, \eol
\beta_{n\Sig^-}^{4 + i5} &=
-\frac 1{24}(35 D^3 + 23 D^2 F + 9 D + 33 D F^2 - 9 F - 123
F^3)\fpi\eolnn
&\qquad - \frac 1{12}
(31 D^3 - 53 D^2 F + 9 D + 57 D F^2 - 9 F - 51 F^3)\fk\eolnn
&\qquad - \frac 1{24}
(D - F)(11 D^2 - 6 D F + 27 F^2 + 9) \feta, \eol
\label\lastcorr
\beta_{\Sig^0\Xi^-}^{4 + i5} &=
-\frac 1{24\sqrt 2}
(35 D^3 - 23 D^2 F + 9 D + 33 D F^2 + 9 F + 123 F^3) \fpi\eolnn
&\qquad -\frac 1{12\sqrt 2}
(31 D^3 + 53 D^2 F + 9 D + 57 D F^2 + 9 F + 51 F^3) \fk\eolnn
&\qquad - \frac 1{24\sqrt 2}
(D + F)(11 D^2 + 6 D F + 27 F^2 + 9)\feta.\eeol
\eeq
Here $\mu$ is an arbitrary renormalization scale.
The $\mu$ dependence of these results is cancelled by the $\mu$
dependence of $O(m_s)$ terms in the effective lagrangian such as
\eq
\label\higher
\frac {c(\mu)}{\Lam}\, \tr\left( \mybar B M s \cdot A B \right).
\eeq
If we take $\mu \simeq \Lam$, there are no large logarithms in the
higher order coefficients, and the correction is dominated by the
logarithmically enhanced terms (computed above) near the chiral
limit.
In the real world these logarithms are not very large, but we expect
that the logarithmic terms will give a good indication of the actual
size of the corrections.

In the $SU(3)$ limit, using (24)--(30) we find
\eq
\label\exact
g_{ab}^c(0) = D' d_{ab}^c + F' f_{ab}^c,
\eeq
where
\eq
\label\dpfp
\eqalign{
D' &= D - \frac 32 D (3 D^2 + 5 F^2 + 1)\,
\frac{m^2}{16\pi^2 f^2} \ln\frac{m^2}{\mu^2}, \cr
F' &= F - \frac 16 F (25 D^2 + 63 F^2 + 9)\,
\frac{m^2}{16\pi^2 f^2} \ln\frac{m^2}{\mu^2}, \cr}
\eeq
and $m$ is the common meson mass.

This shows that for purposes of evaluating $SU(3)$ breaking in
semileptonic hyperon decays, it is misleading to present the
results in terms of $D$ and $F$ defined in the effective lagrangian
eq.\ \Beff, since large corrections to $D$ and $F$ do not necessarily
correspond to large $SU(3)$ breaking.
We therefore consider an expansion in $D'$ and $F'$, where $m$ is chosen
to be some appropriate average meson mass (see below) treated as $O(m_s)$
for purposes of power counting.
This expansion can easily be made well-defined to all orders, for example
by defining the relations eq.\ \exact\ to be exact in the limit where all
mesons have a common mass $m$.

The parameter $m$ in eq.\ \dpfp\ is a redundant parameter in this expansion
analogous to the renormalization scale $\mu$ in conventional perturbation
theory.
In a world where the quark mass differences are small compared
to the average quark mass, it is clear that $m$ should be chosen to be close
to the average meson mass.
In our world, $SU(3)$-breaking quark mass differences are of order $m_s$,
and it is not clear {\it \'a priori} how to choose $m$.
We simply choose $m$ in order to minimize the corrections to the
lowest-order results.
This choice is justified {\it \'a fortiori} by the fact that we obtain
a reasonable value for $m$ ($\simeq 300 \MeV$), and by the fact that
the corrections expressed in terms of $D'$ and $F'$ are small.
This is a non-trivial feature of the logarithmically-enhanced corrections,
since both $SU(3)$ and chiral symmetry breaking are controlled by the same
parameter, namely $m_s$.

The large corrections to the lowest-order results in terms of $D$ and $F$
indicate that chiral perturbation theory is breaking down for this process.
However, we wish to emphasize that this breakdown of chiral perturbation
theory does not necessarily imply a breakdown of the expansion in terms
of $D'$ and $F'$.
In \Beff, $D$ and $F$ have an absolute physical significance in terms of the
couplings of the light mesons to baryons in chiral perturbation theory.
In contrast our parameters $D'$ and $F'$ are defined through $SU(3)$
relations.

In order to determine $D'$ and $F'$ we performed a fit to the decay rates
and asymmetry data quoted by the Particle Data group \PDG.
Because we expect that higher-order terms in the chiral expansion
give corrections of order
\eq
\frac{m_K^2}{16\pi^2 f^2} \sim 0.25,
\eeq
we have increased the uncertainties on the measured values of $g_1$ by $20\%$.
(More information about our fit is presented in appendix A.)
Fitting to the lowest-order results gives
\eq
\label\lowfit
D = 0.85\pm 0.06, \qquad
F = 0.52\pm 0.04,
\eeq
with $\chi^2 = 6.1$ for 9 degrees of freedom.
(Recall that $D' = D$, $F' = F$ at lowest order.)

Using $m = 260 \MeV$ and $\mu = m_\rho$, the corrections to $g_1(0)$
for all decay modes are less than $20\%$, and we obtain the best-fit values
\eq
\label\corrfit
D' = 0.87\pm 0.06, \qquad
F' = 0.53\pm 0.04,
\eeq
with $\chi^2 = 6.3$.

\section{$SU(3)$ Breaking and the ``EMC Effect''}

$SU(3)$ breaking is important for determining the value
of various strange-quark matrix elements of nucleons.
In this section, we briefly present the predictions of the expansion
discussed in section 3 to the extraction of the matrix element
\eq
\Delta s(Q^2) \equiv
\bra{p, s} \left. \mybar s \gamma_\mu \gamma_5 s \right|_{Q^2}
\ket{p, s},
\eeq
where $\ket{p, s}$ is a proton state with spin $s$.
The unexpectedly large value of this matrix element extracted from
analysis of EMC data \EMC\ is often called the ``EMC effect'' and has
attracted a good deal of attention in the theoretical literature
\muchado.

Combining a (rigorous, QCD-derived) sum rule with isospin invariance
allows us to derive the relation
\eq
\eqalign{
\int_0^1 dx\, g_1(x, Q^2) &=
\frac 1{36}\left[ 3 g_A + 5 (\Delta u + \Delta d - 2\Delta s)
+ 12 \Delta s(Q^2) \right] \cr
& \qquad\quad \times \left[ 1 - \frac{\al_{\rm s}(Q^2)}{\pi}
+ O(\al_{\rm s}^2) \right]
+ O(\Lam^2 / Q^2), \cr}
\eeq
where $g_A \simeq 1.25$ is the nucleon axial coupling.
The left-hand side extracted (with extrapolation) from the EMC data is
$0.126\pm 0.018$ \EMC, where we have added systematic and statistical
errors in quadrature.
We have
\eq
\Del u + \Del d - 2\Del s = (3F - D)
\left[1 + \frac 1{16\pi^2 f^2} \gamma \right],
\eeq
where
\eq
\label\emcorr
\eqalign{
\gamma & = 3 (D + F)^2 \fpi
- \frac 16 (9 + 7 D^2 - 18 DF + 27 F^2) \fk \cr
&\qquad\quad
- \frac 13 (D - 3 F)^2 \feta. \cr}
\eeq
Expressing the results in terms of $D'$ and $F'$ and using our
best-fit values, we obtain
\eq
\label\emceffect
\eqalign{
\Del s &= (-0.13 \pm 0.07) M_p,\cr
\Del u + \Del d + \Del s &= (\phantom{-}0.12 \pm 0.19) M_p,\cr}
\eeq
whereas we obtain $\Del s = (-0.20 \pm 0.06) M_p$ and
$\Del u + \Del d + \Del s = (0.06 \pm 0.18) M_p$
if we do not include $SU(3)$-breaking corrections.

We may not trust the predicted $SU(3)$ breaking in
eq.\ \emcorr\ quantitatively, since $O(m_s)$ corrections are not
included.
However, it is worth noting that the corrections we have computed
significantly reduce the value of $\Delta s$, suggesting that the
``EMC effect'' may be less striking than commonly thought.

\section{Nonleptonic Decays}

In this section, we consider nonleptonic decays as another application
of the formalism discussed in section 3.
We will find that our results tell much the same story as the semileptonic
decays: there are large corrections to the lowest-order predictions of
chiral symmetry, but corrections to $SU(3)$ relations are $\lsim 10\%$.

We consider only the predictions for the $s$-wave nonleptonic decay
amplitudes here, since the chiral perturbation theory predictions
for the $p$-wave amplitudes do not follow from $SU(3)$ alone.
The effective $\Delta S = 1$ lagrangian at the weak scale can be
written
\eq
\scr L_{\Del S = 1} = \frac{4G_F}{\sqrt 2} V_{ud} V^*_{us}
(\mybar q_L \gamma^\mu S_1 q_L) (\mybar q_L \gamma_\mu S_2 q_L),
\eeq
where
\eq
q = \pmatrix{u \cr d \cr s \cr}, \quad
S_1 = \pmatrix{0 & 0 & 1 \cr 0 & 0 & 0 \cr 0 & 0 & 0 \cr}, \quad
S_2 = \pmatrix{0 & 0 & 0 \cr 1 & 0 & 0 \cr 0 & 0 & 0 \cr}.
\eeq
We follow standard practice and assume the dominance of the
$\Delta I = {1\over 2}$ amplitudes.
We therefore add to the effective lagrangian the terms
\eq
\label\nonleptoniclagrangian
\del\scr L_{\Del S = 1} = h_D \tr(\mybar B \{H, B\})
+ h_F \tr(\mybar B [H, B]),
\eeq
where
\eq
H \equiv \xi^\dagger S_2 S_1 \xi \mapsto U H U^\dagger.
\eeq
Previous authors \Jnon\ have also included a term which is higher order
in the derivative expansion on the grounds that its coefficient, $h_\pi$,
as measured in $\Delta S=1$ kaon decays is larger than expected by
dimensional analysis.
We choose to work to a consistent order in the chiral expansion and will
neglect this term.  The enhancement of $h_\pi$ is attributed to the
$\Delta I={1\over 2}$ rule which may be violated in these decays (see
below), making special treatment of this term somewhat suspect.
Also, we have no information about other higher order terms which could
also have anomalously large coefficients.
In any case, we are interested primarily in the question of the size of
$SU(3)$ violation, and barring accidental cancellations, we expect that
the logarithmically-enhanced corrections to give a good indication of
the size of the corrections.

The $s$-wave decay amplitude for $B_a\goto B_b\pi$ can be written as
\eq
\scr M_s = G_F m_\pi^2\ \bar{u}_a \scr A_{ab} u_b,
\eeq
where ${\cal A}_{ab}$ is the dimensionless $s$-wave (parity violating)
amplitude as defined in ref.\ \PDG.

Assuming the $\Delta I={1\over 2}$ rule, there are three isospin relations
among the seven decay amplitudes that have been measured:
\eq
\label\treenl
\eqalign{
\sqrt 2 \scr A(\Sig^+ \goto p\pi^0) - \scr A(\Sig^+ \goto n\pi^+)
+ \scr A(\Sig^- \goto n\pi^-) &= 0,
\qquad (5\%) \cr
\scr A(\Lam \goto p \pi^0) + \sqrt 2 \scr A(\Lam \goto n \pi^0) &= 0,
\qquad (1\%) \cr
\scr A(\Xi^- \goto \Lam \pi^-) + \sqrt 2 \scr A(\Xi^0 \goto \Lam \pi^0)
&= 0. \qquad\, (8\%) \cr}
\eeq
The experimental deviation from these relations is shown in parentheses.
(Details on the data and fits can be found in appendix B.)

These ``isospin'' relations do not work significantly better
than the $SU(3)$ relations (see below), suggesting that the
$\Del I = \frac 12$ rule may not be accurate for these decays.
However, since we are interested primarily in the size of $SU(3)$
violation, it is sufficient to assume the $\Del I = \frac 12$ form
eq.\ \nonleptoniclagrangian\ for the lagrangian.

The predictions for the remaining independent $s$-wave amplitudes are
\eq
\scr A_{ab} = \al_{ab} \left[
1 + \frac 1{16\pi^2 f^2} (\beta_{ab} + \epsilon) \right],
\eeq
where $\al$ is the lowest-order prediction, and $\beta$ and $\epsilon$ are
the corrections.
$\epsilon$ contains pion wavefunction renormalization and renormalization
of $f_\pi$.
These effects are the same for all decays, and therefore do not affect
the $SU(3)$ predictions.
We will not need the explicit expressions for these corrections.

The tree-level results for the four independent amplitudes are
\eq
\eqalign{
\al_{n \Sig^+} &= 0, \cr
\al_{n \Sig^-} &= -h_D + h_F, \cr
\al_{p \Lam} &= \frac 1{\sqrt 6} (h_D + 3 h_F), \cr
\al_{\Lam \Xi^-} &= \frac 1{\sqrt 6} (h_D - 3 h_F), \cr}
\eeq

At lowest order we can eliminate $h_D$ and $h_F$ to obtain an $SU(3)$
relation among the three non-vanishing amplitudes: the Lee--Sugawara relation
\eq
\Del_{\rm LS} \equiv
\frac 3{\sqrt 6} \scr A_{n\Sigma^-} + \scr A_{p\Lambda} +
2 \scr A_{\Lambda\Xi^-} = 0.
\eeq
(This relation is often written including a term proportional to
$\scr A_{n\Sig^+}$.)

The leading chiral corrections are
\eqa
\label\firstnlcor
\beta_{n\Sig^+} &= 0, \eol
\beta_{n \Sig^-} &= \frac 1{24} \Bigl[7(h_F-h_D)
+ h_D(51D^2-6DF+27F^2) \eolnn
&\qquad\qquad\qquad - h_F(3D^2-54DF+27F^2)\Bigr] \fpi\eolnn
&\qquad + \frac 1{12} \Bigl[5(h_D-h_F) + h_D(39D^2-30DF+27F^2) \eolnn
&\qquad\qquad\qquad\quad - h_F(15D^2-54DF+27F^2)\Bigr]\fk\eolnn
&\qquad + \frac 3{8}(h_D-h_F)(1+3D^2-6DF+3F^2)\feta, \eol
\beta_{p \Lam} &= \frac 1{24\sqrt 6} \Bigl[7(h_D+3h_F)
- h_D(171D^2-162DF+27F^2) \eolnn
&\qquad\qquad\qquad -h_F(81D^2+54DF+81F^2) \Bigr]\fpi\eolnn
&\qquad + \frac 1{12\sqrt 6}\Bigl[ -5(h_D+3h_F)
+ h_D(9D^2-90DF-27F^2) \eolnn
&\qquad\qquad\qquad\quad -h_F(45D^2+54DF+81F^2) \Bigr]\fk\eolnn
&\qquad - \frac 1{8\sqrt 6} (h_D+3h_F)(3+D^2+6DF+9F^2)\feta, \eeol
\eeq
\eqa
\label\lastnlcor
\beta_{\Lam \Xi^-} &= \frac 1{24\sqrt 6} \Bigl[7(h_D-3h_F)
- h_D(171D^2-162DF+27F^2) \eolnn
&\qquad\qquad\qquad + h_F(81D^2-54DF+81F^2) \Bigr]\fpi\eolnn
&\qquad + \frac 1{12\sqrt 6}\Bigl[ -5(h_D-3h_F) +
h_D(9D^2+90DF-27F^2) \eolnn
&\qquad\qquad\qquad\quad +h_F(45D^2-54DF+81F^2) \Bigr]\fk\eolnn
&\qquad - \frac 1{8\sqrt 6} (h_D-3h_F)(3+D^2-6DF+9F^2)\feta. \eol
\eeq
Defining $h'_D$ and $h'_F$ in analogy to $D'$ and $F'$ we obtain
from \firstnlcor--\lastnlcor
\eq
\eqalign{
h'_D &= h_D - {1\over 2} \Bigl[ h_D(1+13D^2+9F^2)+18h_FDF \Bigr]
{m^2\over16\pi^2f^2}\ln{m^2\over\mu^2},\cr
h'_F &= h_F - {1\over 2} \Bigl[ h_F(1+ 5D^2+9F^2)+10h_DDF \Bigr]
{m^2\over16\pi^2f^2}\ln{m^2\over\mu^2}. \cr}
\eeq

A fit to the data using the lowest-order predictions gives
\eq
h_D = -0.55 \pm 0.32, \qquad
h_F = 1.37 \pm 0.17,
\eeq
with $\chi^2=0.06$ for 1 degree of freedom.
To account for the theoretical error due to $O(m_s)$ terms in the
expansion we have again added $20\%$ in quadrature to the experimental
errors before doing the fit.
With only one degree of freedom the errors quoted should be taken as
indicative only, but it is clear that the lowest-order predictions
fit the data well.

Expressing our results in terms of $h'_D$ and $h'_F$ and taking
$\mu = m_\rho$ and $m = 320 \MeV$, we find that all the
logarithmically-enhanced $SU(3)$ corrections are less than $10\%$ and the
fit still works well:
\eq
h'_D = -0.56 \pm 0.40, \qquad
h'_F = 1.31 \pm 0.18,
\eeq
with $\chi^2 = 0.30$.
Thus there is every indication that the $SU(3)$ expansion is well-behaved.
This is to be contrasted to the {\it chiral} expansion, in which
corrections to the individual decay amplitudes are $\sim 50\%$.

Chiral symmetry also gives a prediction for the $p$-wave decay amplitudes
which does not follow from $SU(3)$ alone.
These predictions do not work well \Jnon, supporting our conclusion that
$SU(3)$ may be a better symmetry than chiral symmetry.

Including the corrections for the best fit $h'_D$ and $h'_F$ the
Lee--Sugawara relation becomes
\eq
\Del_{\rm LS} = 0.29 \pm 0.13,
\eeq
which is to be compared with the experimental value of $-0.23 \pm 0.03$.
The expected size of the $O(m_s)$ contributions is $\sim 0.4$,
so the fact that the predicted sign of $\Del_{\rm LS}$ is wrong
does not imply that our expansion is breaking down.

\section{Conclusions}

We have investigated the question of $SU(3)$ breaking for weak
hyperon decays in the context of chiral perturbation theory.
One major difference between our work and previous work is that we have
emphasized that large explicit chiral symmetry breaking does not
necessarily imply large $SU(3)$ breaking.
We have found that $SU(3)$ breaking is less than $20\%$, which is
what is expected on the basis of dimensional analysis.
Although we cannot conclude from our analysis that the expansion is under
control, there is no sign that it is breaking down, unlike the usual
chiral expansion.

We also used this expansion to analyze the ``EMC effect,'' and showed
that the $SU(3)$-breaking corrections reduce the extracted value of
the matrix element $\bra p \mybar s \gamma_\mu \gamma_5 s \ket p$
by $35\%$.

\appendix{A}{Fit to Semileptonic Decays}

In this appendix, we present some details of the fit to semileptonic
hyperon decays used in this paper.
We use both decay rate and asymmetry data taken from the most recent
Particle Data Group (PDG) compilation \PDG.
For the asymmetry data, we directly use the average values for
$g_A / g_V$ quoted by the PDG.
To convert the decay rates into values for $g_1$, we keep the full
kinematic dependence on the baryon masses, since these effects turn
out to be numerically important.
The data we use is displayed in table 1.

\def \phm {\phantom{-}}
\vbox{ \vskip 20pt \centerline{
\vbox{ \offinterlineskip
\halign {\vrule#& \hfil#\hfil& \vrule#& \hfil#\hfil& \vrule#&
\hfil#\hfil& \vrule# \cr
\noalign{\hrule}
height2pt& \omit&& \omit&& \omit& \cr
&\qquad\qquad\qquad&& \qquad lifetime\qquad&& \qquad asymmetry\qquad&\cr
height2pt&\omit&&\omit&&\omit&\cr \noalign{\hrule}
height2pt&\omit&&\omit&&\omit&\cr
&$n^{\phm}\goto p^{\phm}$&&
$\phm1.323 \pm 0.003$ && $\phm1.257\pm0.003$ &\cr
height2pt&\omit&&\omit&&\omit&\cr \noalign{\hrule}
height2pt&\omit&&\omit&&\omit&\cr
&$\Sig^-\goto\Lam^{\phm}$&& $\phm0.609\pm0.029$ && $\phm0.62\pm0.44$ &\cr
height2pt&\omit&&\omit&&\omit&\cr \noalign{\hrule}
height2pt&\omit&&\omit&&\omit&\cr
&$\Lam^{\phm}\goto p^{\phm}$&& $-0.972\pm0.018$ && $-0.879\pm0.021$ &\cr
height2pt&\omit&&\omit&&\omit&\cr \noalign{\hrule}
height2pt&\omit&&\omit&&\omit&\cr
&$\Sig^-\goto n^{\phm}$&& $\phm0.442\pm0.021$ && $\phm0.340\pm0.017$ &\cr
height2pt&\omit&&\omit&&\omit&\cr \noalign{\hrule}
height2pt&\omit&&\omit&&\omit&\cr
&$\Xi^- \goto\Sig^0$&& $\phm0.96\phantom{0}\pm0.19\phantom{0}$ && ------ &\cr
height2pt&\omit&&\omit&&\omit&\cr \noalign{\hrule}
height2pt&\omit&&\omit&&\omit&\cr
&$\Xi^- \goto \Lam^{\phm}$&& $\phm0.473\pm0.026$ && $\phm0.306\pm0.061$ &\cr
height2pt&\omit&&\omit&&\omit&\cr \noalign{\hrule} }} } \vskip 10pt }
\centerline{Table 1: Values for $g_1(0)$ extracted from 1992 PDG}

The decay rate and asymmetry determinations of $g_1$ are inconsistent
if we assume only the errors quoted by the PDG.
This is either a symptom of systematic errors in the experiments or an
indication that higher-order corrections are important.
We expect that higher order terms in the chiral expansion will give
rise to $\sim 20\%$ corrections, and so we added this amount in quadrature
to all the quoted errors to take into account the theoretical uncertainty.
When we do this, all the errors on all determinations have a
sizable overlap, and reasonable fits are obtained (see the text).

\appendix{B}{Fit to Nonleptonic Decays}

In this appendix, we give some details about the data used to fit
the $s$-wave nonleptonic decay amplitudes.
The decays have $s$- and $p$-wave components with a possible relative
phase, and so in principle three pieces of information are required to
extract the $s$-wave amplitudes.
We used the total lifetime and the asymmetry parameter $\al$ quoted in
the 1992 PDG \PDG, and neglected final-state phase shifts.
This is consistent since final-state phase shifts are higher order in
the $SU(3)$ expansion.
Table 2 shows the amplitudes obtained in this way.

\vbox{ \vskip 20pt \centerline{
\vbox{ \offinterlineskip
\halign {\vrule#& \hfil#\hfil& \vrule#& \hfil#\hfil& \vrule#\cr
\noalign{\hrule}
height2pt&\omit&&\omit& \cr
&\qquad Decay\qquad&&\qquad\quad ${\cal A}$\quad\qquad& \cr
height2pt&\omit&&\omit& \cr\noalign{\hrule}
height2pt&\omit&&\omit& \cr
& $\Lambda^{\phm}\goto p^{\phm}$&& $\phm1.43\pm 0.01$& \cr
height2pt&\omit&&\omit& \cr\noalign{\hrule}
height2pt&\omit&&\omit& \cr
& $\Lambda^{\phm}\goto n^{\phm}$&& $\phm1.04\pm 0.01$& \cr
height2pt&\omit&&\omit& \cr\noalign{\hrule}
height2pt&\omit&&\omit& \cr
& $\Sigma^+\goto n^{\phm}$&& $\phm0.06\pm 0.01$& \cr
height2pt&\omit&&\omit& \cr\noalign{\hrule}
height2pt&\omit&&\omit& \cr
& $\Sigma^+\goto p^{\phm}$&& $\phm1.44\pm 0.05$& \cr
height2pt&\omit&&\omit& \cr\noalign{\hrule}
height2pt&\omit&&\omit& \cr
& $\Sigma^-\goto n^{\phm}$&& $\phm1.88\pm 0.01$& \cr
height2pt&\omit&&\omit& \cr\noalign{\hrule}
height2pt&\omit&&\omit& \cr
& $\Xi^0\goto\Lambda^{\phm}$&& $\phm1.51\pm 0.01$& \cr
height2pt&\omit&&\omit& \cr\noalign{\hrule}
height2pt&\omit&&\omit& \cr
& $\Xi^-\goto\Lambda^{\phm}$&& $-1.98\pm 0.01$& \cr
height2pt&\omit&&\omit& \cr\noalign{\hrule} }}   } \vskip 10pt}
\centerline{Table 2: Values for $s$-wave amplitude
${\cal A}$ from 1992 PDG}

Just as for the semileptonic decay amplitudes, we increased the errors on
the $\scr A$ by $20\%$ to account for the theoretical uncertainty arising
from $O(m_s)$ corrections.
When this is done, the data is consistent, and good fits are obtained
(see text).

\section{Acknowledgements}

We would like to thank M. Suzuki for discussions.
This work was supported by the Director, Office of Energy Research, Office
of High Energy and Nuclear Physics, Division of High Energy Physics of the
U.S.\ Department of Energy under Contract DE-AC03-76SF00098.

\listrefs
\bye